# Generating Pareto optimal dose distributions for radiation therapy treatment planning

Dan Nguyen, Azar Sadeghnejad Barkousaraie, Chenyang Shen, Xun Jia, and Steve Jiang

Medical Artificial Intelligence and Automation (MAIA) Laboratory,
Department of Radiation Oncology, UT Southwestern Medical Center, Dallas, TX, USA
`Dan.Nguyen@UTSouthwestern.edu`

**Abstract.** Radiotherapy treatment planning currently requires many trial-and-error iterations between the planner and treatment planning system, as well as between the planner and physician for discussion/consultation. The physician's preferences for a particular patient cannot be easily quantified and precisely conveyed to the planner. In this study we present a real-time volumetric Pareto surface dose generation deep learning neural network that can be used after segmentation by the physician, adding a tangible and quantifiable endpoint to portray to the planner. From 70 prostate patients, we first generated 84,000 intensity modulated radiation therapy plans (1,200 plans per patient) sampling the Pareto surface, representing various tradeoffs between the planning target volume (PTV) and the organs-at-risk (OAR), including bladder, rectum, left femur, right femur, and body. We divided the data to 10 test patients and 60 training/validation patients. We then trained a hierarchically densely connected convolutional U-net (HD U-net), to take the PTV and avoidance map representing OARs masks and weights, and predict the optimized plan. The HD U-net is capable of accurately predicting the 3D Pareto optimal dose distributions, with average [mean, max] dose errors of [3.4%, 7.7%](PTV), [1.6%, 5.6%](bladder), [3.7%, 4.2%](rectum), [3.2%, 8.0%](left femur), [2.9%, 7.7%](right femur), and [0.04%, 5.4%](body) of the prescription dose. The PTV dose coverage prediction was also very similar, with errors of 1.3% (D98) and 2.0% (D99). Homogeneity was also similar, differing by 0.06 on average. The neural network can predict the dose within 1.7 seconds. Clinically, the optimization and dose calculation is much slower, taking 5-10 minutes.

## 1 Introduction

Radiation therapy is one of the major cancer therapy modalities, accounting for two-thirds of cancer patients in the US, either standalone or in conjunction with surgery, chemotherapy, immunotherapy, etc. In the typical current treatment planning workflow, a treatment planner interacts with a commercial treatment planning system to solve an inverse optimization problem, either in an intensity modulated radiation ther-apy (IMRT)[1-3] or volumetric modulated arc therapy (VMAT)[4-7] setting. The planner manually tunes many hyperparameters, such as dose-volume constraints and weightings, to control the tradeoff between multiple clinical objectives. These hyperparameters are meticulously tuned in a time-consuming trial-and-error fashion to reach a suitable clinical solution. In addition, many rounds feedback from the physician is needed for the physician to discuss the plan quality with the planner and to properly portray their desired tradeoffs. This is largely due to the fact that the physician's preferences for a particular patient cannot be fully quantified and precisely conveyed to the planner. This trial-and-error process results in hours of planning time, and the many iterations of physician feedback may extend the time to several days until the plan is accepted.

Recently, deep learning with multi-layered neural networks has exploded in progress, particularly in computer vision. We realize that these new developments can be utilized to solve aspects of the treatment planning problem. Specifically, deep learning can be utilized to quickly realize the physician's preferences in a tangible and quantifiable manner that can be presented to the treatment planner prior to treatment planning. In this study we present a real-time Pareto surface dose generation deep learning neural network that can be used immediately after segmentation by the physician. Pareto optimal plans are the solutions to a multicriteria problem with various tradeoffs. In particular, the tradeoff lies with the dose coverage of the tumor and the dose sparing of the various critical structures. The benefit of the of such a model is two-fold. First, the physician can interact with the model to immediately view a dose distribution, and then adjust some parameters to push the dose towards their desired tradeoff in real time. This also allows for the physician to quickly comprehend the kinds of the tradeoffs that are feasible for the patient. Second, the treatment planner, upon receiving the physician's desired dose distribution, can quickly generate a fully deliverable plan that matches this dose distribution, saving time in tuning the optimization hyperparameters and discussing with the physician. We developed, trained, and tested the feasibility of the model on prostate cancer patients planned with 7 beam IMRT.

## 2 Methods

### 2.1 Prostate patient data and Pareto plan generation

We acquired the anatomical data for 70 prostate patients, in terms of the segmentation of the planning target volume (PTV) and the organs-at-risk, including bladder, rectum, left femur, right femur, and body. Ring and skin structures were added as tuning structures. The patient contours and dose arrays were formatted into 192 x 192 x 64 arrays at 2.5 mm$^3$ voxel size. We then calculated the dose influence arrays for these 70 patients, for a 7 equidistant coplanar beam plan IMRT, with 2.5 mm$^2$ beamlets at 100 cm isocenter—a typical setup for prostate IMRT. Using this dose calculation data, we generated IMRT plans that sampled the Pareto surface, representing various tradeoffs between the PTV and OARs. The multicriteria objective can be written as

$$\begin{aligned}&\underset{x}{\text{minmize}} && \{f_{PTV}(x), f_{OAR_1}(x), f_{OAR_2}(x), \ldots, f_{OAR_n}(x)\}\\ &\text{subject to} && x \geq 0,\end{aligned} \quad (1)$$

where $x$ is the fluence map intensities to be optimized. There exists individual objectives, $f_s(x) \; \forall s \in PTV, OAR$, for the PTV and each of the OARs. Typically, the objective function is designed such that the goal is to deliver the prescribed dose to the PTV, while minimizing the dose to each OAR. Due to the physical aspects of external beam radiation, it is impossible to give the PTV exactly the prescription dose without irradiating normal tissue. Thus, we arrive at a multicriteria objective, where there does not exist a single optimal $x^*$ that would minimize all $f_s(x) \; \forall s \in PTV, OAR$. For a proof of concept in this study, we choose to use the L2-norm to represent the objective, $f_s(x) = \frac{1}{2}\|A_s x - p_s\|_2^2$. Here, $A_s$ is the dose influence matrix for a given structure, and $p_s$ is the desired dose for a given structure, assigned as the prescription dose if $s$ is the PTV, and 0 otherwise. This allows for us to linearly scalarize[8] the multicritera optimization problem into a single-objective, convex optimization problem,

$$\begin{aligned}&\underset{x}{\text{minmize}} && \frac{1}{2}\sum_{s \in S} w_s^2 \|D_s x - p_s\|_2^2\\ &\text{subject to} && x \geq 0.\end{aligned} \quad (2)$$

The key to scalarizing the problem is the addition of $w_s$, which are the tradeoff weights for each objective function, $f_s(x) \; \forall s \in PTV, OAR$. With different values of $w_s$, different Pareto optimal solutions are generated. Using an in-house GPU-based proximal-class first-order primal-dual algorithm, Chambolle-Pock[9], we generated 1,200 pseudo-random plans per patient, totaling to 84,000 plans.

**Table 1.** Weight assignment categories. The function $rand(lb, ub)$ represents a uniform random number between a lower bound, $lb$, and an upper bound, $ub$.

| Category | Description |
|---|---|
| Single organ spare | $w_{s_i} = rand(0,1)$ <br> $w_{OAR \setminus s_i} = rand(0,0.1)$ |
| High weights | $w_s = rand(0,1) \; \forall s \in OAR$ |
| Medium weights | $w_s = rand(0,0.5) \; \forall s \in OAR$ |
| Low weights | $w_s = rand(0,0.1) \; \forall s \in OAR$ |
| Extra low weights | $w_s = rand(0,0.05) \; \forall s \in OAR$ |
| Controlled weights | $w_{bladder} = rand(0,0.2)$ <br> $w_{rectum} = rand(0,0.2)$ <br> $w_{lt\,fem\,head} = rand(0,0.1)$ <br> $w_{rt\,fem\,head} = rand(0,0.1)$ <br> $w_{shell} = rand(0,0.1)$ <br> $w_{skin} = rand(0,0.3)$ |

The generation of each plan entailed assigning pseudo-random weights to the organs-at-risk. The weight for the PTV was kept at 1. The weight assignment fell into 1 of 6 categories as shown in Table 1. For each patient, 100 plans for each organ-at-risk used the single organ spare category (bladder, rectum, left femoral head, right femoral head, shell, skin), totaling to 600 single organ spare plans for each patient. To ensure a larger sampling of weights, another 100 plans of the high, medium, low, and extra low weights were generated, as well as 200 plans of the controlled weights category. The bounds for the controlled weights were chosen through trial and error such that the final plan generated had a high likelihood of being in acceptable clinical bounds for an inexperienced human operator, but not necessarily acceptable for an experienced physician. In total 1,200 plans were generated per patient. With 70 patients, the total number of plans generated was 84,000 plans.

## 2.2 Deep learning architecture

We utilized a volumetric Hierarchically Dense U-net (HD U-net) architecture[10], as shown in Figure 1, which adds in the densely connected convolutional layers[11] into the U-net architecture[12]. The HD U-net was trained to take as input the PTV contour, the body contour, and an avoidance map representing OARs masks assigned their respective $w_s$, and to predict the optimized 3D dose distribution.

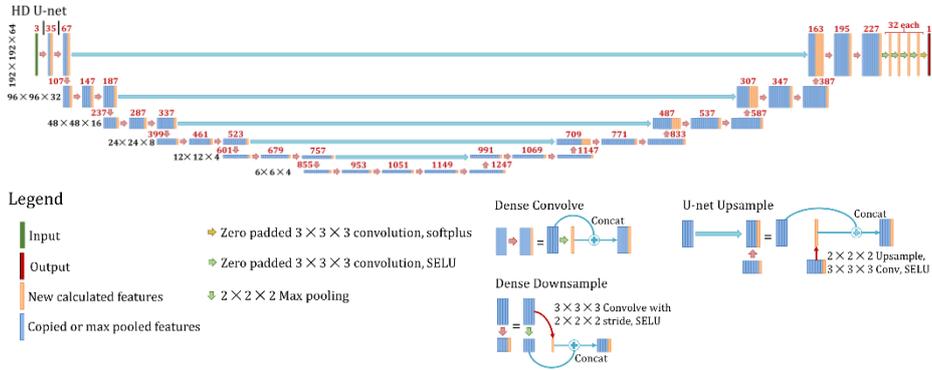

**Fig 1.** Specific HD U-net architecture used in this study. Black numbers to the left of the model represents the current dimensions of the 3D data at each hierarchy. The red numbers indicate the number of feature maps present at the current layer in the neural network. The large number of features maps are due to the densely connected convolutional layers.

Specifically, our HD U-net architecture has 5 max pooling and 5 upsampling operations, ultimately reducing our image size from 192 x 192 x 64 voxels to 6 x 6 x 4 voxels (the lowest level max pooling/upsampling layer reduces/expands leaves the slice dimension untouched), and back to 192 x 192 x 64 voxels. Skip connections are added between the first half and second half of the network, to allow for the propagation of local information with the global information. Densely connected convolutional connections are added in each block of the network, allowing for efficient infor-mation flow of features. The non-linearity used after each convolution was the scaled exponential linear unit (SELU) as presented by Klambauer et al. for self-normalizing neural networks[13]. The study proved, using the Banach fixed-point theorem, that by having the SELU nonlinear activation, the neuron activations automatically converge towards zero mean and unit variance. Also, by the paper suggestion, we did not include batch normalization, as that disrupts the self-normalizing property of SELU-based networks. Since the densely connection convolutional layers allows for less trainable parameters to be used, instead of doubling the number of kernels after every max pooling, we increased number of kernels by 1.25 fold, to the nearest integer. We chose our final activation layer as the softplus activation, as our output data is non-negative and we had found that it is much more stable for training than linear and the rectified linear unit (ReLU) when using SELU as the hidden layer activation.

## 2.3   Training and Evaluation

We randomly divided the data to 10 test patients and 60 model development (training and validation) patients. The 10 test patients were held out during the entire model development phase, and only used during evaluation. Five instances of the model were trained and validated, using 54 training patients and 6 validation patients, according the schematic outlined in Figure 2.

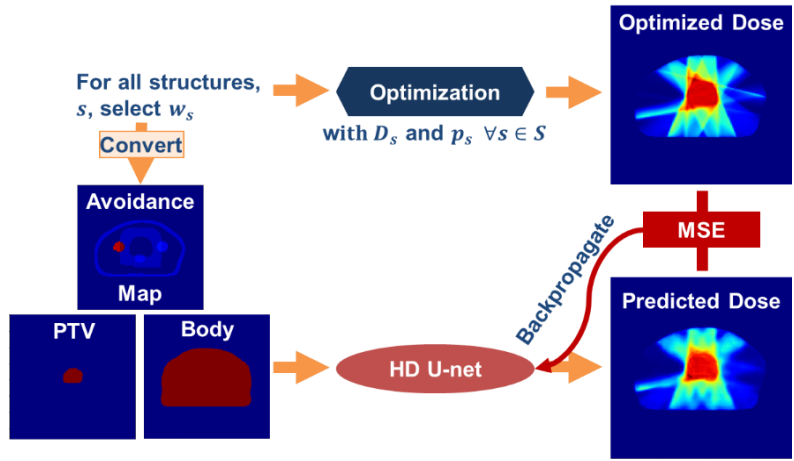

**Fig 2.** Training schematic for the HD U-net architecture

At each training iteration, first a patient is selected and then set of $w_s$ is selected from one of the 1,200 plans. These set of weights are then converted into an avoidance map, which is a single channel of the input that represents the OAR masks assigned their corresponding $w_s$. In addition, the binary mask of the PTV and body are included as input. The HD U-net then makes a prediction using these inputs. The optimized dose, that was generated using the dose influence array and Chambolle-Pock algorithm, is used to minimize against the predicted dose distribution with a mean squared error loss. Alternatively, a plan can be generated on the fly from a given set of $w_s$, but is less efficient for training on a single GPU. During training the model was assessed on the validation data every 200 iterations of training. Each instance of the model used a different set of validation patients for determining the at which iteration the lowest validation score was obtained. Using all 1,200 plans per training patient—64,800 training plans total—we trained the model for 100,000 iterations using the Adam optimizer, with a learning rate of $1 \times 10^{-4}$, using an NVIDIA V100 GPU. The 10 test patients were then evaluated using the trained models.

To equally compare across patients, the test plans were first normalized such that the dose to 95% of the PTV (D95) was equal to the prescription dose. For evaluation criteria, the PTV coverage (D98, D99), PTV max dose (defined as D2 by the ICRU-83 report[14]), homogeneity $\left(\frac{D2-D98}{D50}\right)$, and the structure max and mean doses ($D_{max}$ and $D_{mean}$) were evaluated.

## 3  Results

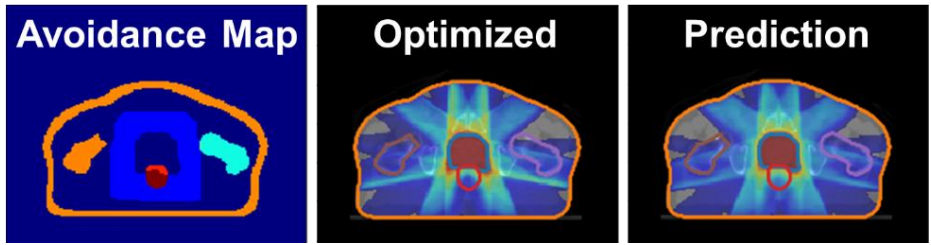

**Fig 3.** Example avoidance map, optimized dose, and prediction for a test patient. The avoidance map represents the structure masks assigned their respective optimization structure weights, $w_s$.

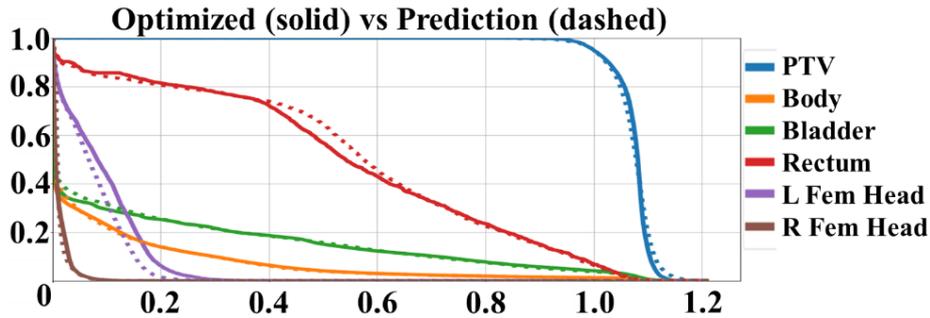

**Fig 4.** Example dose volume histogram for a test patient.

The HD U-net is capable of accurately predicting the Pareto optimal 3D dose distributions, with average mean dose errors of 3.4% (PTV), 1.6% (bladder), 3.7% (rectum), 3.2% (left femur), 2.9% (right femur), and 0.04% (body) of the prescription dose, as compared to the optimized plans. In addition, the HD U-net maintains the average max dose error of 7.7% (PTV), 5.6% (bladder), 4.2% (rectum), 8.0% (left femoral head), 7.7% (right femoral head), and 5.4% (body) of the prescription dose. The PTV dose coverage prediction was also very similar, with errors of 1.3% (D98) and 2.0% (D99) of the prescription dose. On average, the PTV homogeneity between the optimized reference dose and the prediction differed by 0.06. Figure 3 shows the avoidance map, optimized dose and prediction, and Figure 4 shows the dose volume histogram for a test patient.

It took approximately 15 days to train each instance of the model for 100,000 iterations. Figure 5 represents the mean training and validation loss for the HD U-net over the 100,000 iterations of training. The validation curve begins to flatten out at around 80,000 iterations while the training loss continues to decrease. The small standard deviation in validation loss between the model instances indicate the stability and reproducibility of the overall model framework and choice of hyperparameters.

Given any structure weights set and anatomy, the neural network is capable of predicting the dose distribution in 1.7 seconds. Clinically, the optimization and dose calculation for IMRT takes approximately 5-10 minutes to complete. This makes it feasible for the model to be used in a real-time setting with a human operator.

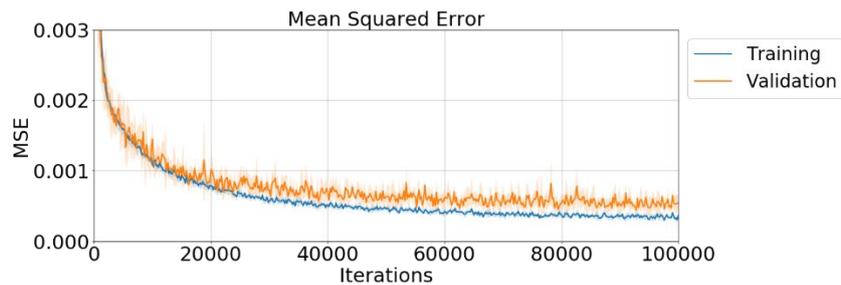

**Fig 5.** Training and validation loss. Solid line represents the mean loss of the 5 model instances trained on different training/validation sets. Error represents one standard deviation.

## 4 Discussion and Conclusion

While other deep learning models designed to learn and predict the dose distribution of a patient plans, based either on historical clinical data or optimized plans to meet standardized clinical criteria, were developed in recent years[10, 15-20], this Pareto dose distribution model, to our knowledge, is the first deep learning model to able to generate any optimized plan from just the anatomy and structure weights. Although the model does not generate the final plan in terms of deliverability, its real-time prediction capabilities allow for it to be used as a tool for the physician quickly generate a dose distribution with realistic tradeoffs between the PTV and various OARs. This can then be given to the planner as an endpoint, alongside the other typical planning information provided by the physician. The treatment planner now has a tangible, physician-preferred endpoint to meet, and the physician gets an initial understanding of what is physically achievable. To further improve the automation, we plan to implement a robust dose mimicking optimization, such as TORA[21], which will automatically generate a deliverable plan given a dose distribution or constraints. We expect that the implementation of such a framework would drastically reduce the number of feedback loops between the planner and physician, and potentially fully automate the treatment planning for simple cases. The valuable time that is saved would allow for the physician and planner to focus on more challenging cases to produce the best achievable plan.